\documentclass[11pt,preprint]{aastex}

\begin{document}
\title{PRIMORDIAL LITHIUM FROM GLOBULAR CLUSTER TURN-OFF STARS: M13 AND M71}

\author{Ann Merchant Boesgaard\altaffilmark{1}}
\affil{Institute for Astronomy, University of Hawai`i at M\-anoa, \\ 
2680 Woodlawn Drive, Honolulu, HI {\ \ }96822 \\ } 

\email{annmb@hawaii.edu}

\author{Constantine~P.~Deliyannis\altaffilmark{1}}
\affil{Department of Astronomy, Indiana University, 727 East 3rd Street, \\
Swain Hall West 319, Bloomington, IN {\ \ }47405-7105 \\ }

\email{cdeliyan@indiana.edu}

\altaffiltext{1}{Visiting Astronomer, W.~M.~Keck Observatory, jointly operated
by the California Institute for Technology and the University of California.}

\begin{abstract}
During Big Bang nucleosynthesis (BBN) in the first 15 minutes of the Universe
some $^7$Li was created along with isotopes of H and He.  The determination of
that primordial value of Li can help constrain the conditions at that time.
The oldest stars with known ages can be found in globular clusters which have
well-determined ages through stellar evolution models.  High-resolution
spectra have been obtained with the Keck I telescope and HIRES of Li in
several unevolved stars in the clusters M13 and M71 with V magnitudes of 17.6
-- 17.9.  Abundances of Li have been determined with spectrum synthesis
techniques and show a range of a factor of 4.  We attribute that spread to
differences in initial angular momentum resulting in different amounts of
spin-down, related mixing, and destruction of Li.  Our results are compared
with similar results for main-sequences and turn-off stars in other globular
clusters.  The range in age of these clusters is 11.2 to 14.2 Gyr for an age
span of 3 Gyr.  These clusters range in [Fe/H] from $-$0.75 to $-$2.24
corresponding to a factor of 30 in metallicity.  The maximum in the Li
abundance for these unevolved stars in all eight clusters is the same
corresponding to Li/H = 3.16 x 10$^{-10}$ while the predicted Li abundance,
based on the deuterium abundance and the BBN predictions, is 5.24 x
10$^{-10}$.
\end{abstract}

\section{INTRODUCTION}

At the very high temperatures and densities during the first 15 minutes of the
Big Bang neutrons and protons combine to form $^2$H, $^3$He, $^4$He, $^7$Li in
what is referred to as Bang nucleosynthesis (BBN).  That original Li can be
destroyed in stars most commonly by fusion with a proton at temperatures of
$\sim$2.5 x 10$^6$ K.  And additional Li can be created in various ways: by
energetic cosmic rays interacting with abundant atoms of He, C, N, O in the
interstellar medium, by runaway thermonuclear reactions in novae explosions,
in supernovae explosions either by the outward shock waves or by neutrino
processes, and in AGB stars through the Cameron-Fowler $^7$Be-transport
mechanism. Models of Galactic chemical evolution explore how such mechanisms
might have raised the Big Bang Li abundance to the present (local) Galactic
value (see Romano et al.~1999).

The best way to find how much primordial Li was created is to look in the
oldest, first-generation stars.  The oldest stars with known ages can be found
in globular clusters where the ages are determined from stellar evolution
models.  The stars with primordial Li would be those first-generation cluster
stars that have not evolved beyond the main-sequence phase of evolution.  Such
stars have faint apparent magnitudes so some studies have looked at Li in
somewhat brighter unevolved turn-off stars.  The most metal-poor star-clusters
are likely to be the oldest and thus potentially have the original Li content
that was made during Big Bang Nucleosynthesis.

A prime example of a globular cluster fitting these criteria is M92.  It has
an age of 13.30 $\pm$0.60 Gyr (Valcin et al.~2020) and [Fe/H] = $-$2.239
$\pm$0.028 (Bailin 2019).  We reported on our Keck/hires results on three
subgiant stars and one turn-off star in M92 in 1995 (Deliyannis, Boesgaard \&
King 1995).  Then in 1998 we discussed the Li results for seven stars near the
turn-off (Boesgaard et al.~1998) where we showed the range in Li of a factor
of three, with values of A(Li) = log N(Li)/N(H) + 12.00 ranging from 2.01 to
2.45 even as specific line strengths of other elements - Fe, Ca, Cr, Ti, Ba -
were similar as seen in their Figure 7.

We have determined Li abundances in five virtually identical turn-off stars in
the 12.7 Gyr old globular cluster M5 (Boesgaard \& Deliyannis 2023) also from
high resolution Keck/HIRES observations.  We found a range in Li abundances of
a factor of 2 in those five stars.  That spread is five times the error in an
individual determination.  The maximum Li abundance in M5 of A(Li) = 2.47 is
similar to that in M92.

Here we add Li in two more two more globular clusters, M13 and M71.  The old
globular cluster, M13, has an age of 13.49 Gyr while M71 is one of the younger
ones at 11.21 Gyr (Valcin et al.~2020).  We found [Fe/H] = $-$0.80 for M71
(Boesgaard et al.~(2005a) and it is one of the more metal-rich globular
clusters in the catalog of Bailin (2019).  That catalog gives [Fe/H] =
$-$1.443 for M13 and $-$0.736 for M71.  Preliminary results for Li in M13 and
M71 were presented at an SPIE meeting (Boesgaard et al.~2000) and at an IAU
symposium (Boesgaard et al.~2005b).

\section{OBSERVATIONS}

In order to find the pristine Li content of globular clusters we need to
observe unevolved stars.  Stellar evolution toward the red giant branch will
reduce the surface abundance of Li due to convective dilution effects as first
shown by Iben (1965, 1967).  This was dilution effect was explored by
Deliyannis, Demarque \& Kawaler (1990) in metal-poor stars.  Such unevolved
stars are faint, however, and long exposures are required.  Our
high-resolution spectra were taken with HIRES on the Keck I 10-m telescope on
Mauna Kea.  Our spectral resolution is $\sim$45,000.  Each night multiple flat
fields were obtained, typically 20, along with some 20 bias frames.
Wavelength calibration spectra of Th-Ar were taken at the beginning and the
end of each night.

For M13 we observed four stars that are virtually identical in their position
in the color-magnitude diagram with V = 17.88 and B-V = 0.51 as seen in Figure
1, left.  Our star selection was based on photometry by Stetson
(priv. comm. 1996); the photometry is now registered in a catalog by Stetson
et al.~(2019).  Spectra of our turn-off stars were obtained on four different
observing nights with individual exposures of 45 minutes.  Some details are
given in Table 1.  The values for the combined signal-to-noise ratios (S/N) in
the Li I region are given in the last column, ranging from 36 to 72.

We could obtain spectra of five turn-off stars in M71 with V = 17.6 -- 17.7.
The positions of these stars in the color-magnitude diagram are shown in
Figure 1, right.  The specifics of the observational data for M71 are given in
Table 2.  There are five stars with very similar values of V and B-V with
spectra taken over seven nights with multiple exposures of typically 45
minutes each.  The combined spectra have S/N values of 48 to 60.

The data reductions for the M71 stars can be found in the paper by Boesgaard
et al.~(2005a); we did not do a Li analysis in that paper.  The procedure for
the M13 turn-off stars was similar to that used for M71 stars.  It involved
IRAF$\footnote{IRAF is distributed by the National Optical Astronomy
Observatories which are operated by NOIRLab, under cooperative agreement with
the National Science Foundation.}$ routines with bias-subtraction,
flat-fielding with division by a normalized flat, and wavelength calibration
for each order from the Th-Ar spectra.  This is similar to the procedures we
used in our papers on Li abundances for M92 (Boesgaard et al.~(1998) and for
M5 (Boesgaard \& Deliyannis 2023).

\section{ABUNDANCE DETERMINATIONS}

Stellar parameters that we used in our paper on abundances in M71 (Boesgaard
et al.~(2005a) are the ones we have used here to find Li abundances.  For M13
stars we used temperatures derived on the scale of Carney (1983), log g = 3.80
and microturbulent velocity, $\xi$, of 1.0 km sec$^{-1}$.  The stellar
parameters that we used are listed for M13 in Table 3 and for M71 in Table 4.
In our previous work on M92 and M5 we have used this Carney temperature scale
and this same analysis method.

We have used the program
MOOG$\footnote{http://www.as.utexas.edu/$~$chris/moog.html}$ Sneden (1973),
Sneden et al.~(2012) to conduct a spectrum synthesis of the Li I region near
6707 \AA.  The synthesized region covers from 6703.354 to 6711.014 and the
linelist has 94 lines, including 39 lines of CN.

Examples of the synthesis fits for two stars in M13 are shown in Figure 2.
Although the stellar temperature of these two stars agree within 10 K, their
Li abundances differ by a factor of 4: A(Li) = 2.45 vs 1.85.  The star with
the weak Li line has a depth of 0.94 while the the line in 38266 has a depth
of 0.82.  The other two stars we observed in M13 are intermediate between
these two values for A(Li).  Values of A(Li) are given in the last column of
Table 3 and span that range from 1.85 to 2.45.

Also given in Table 3 are the measured equivalent widths of the Li I feature
for each star.  We used the ``blends'' feature in MOOG, including the Li
hyperfine lines and an Fe I line at 6707.44, to determine A(Li).  That line
strength and Li determination are also given in Table 3.  Although we prefer
the Li abundances found from the synthesis method, we see that the abundances
are very similar when done by the blends method.  The mean difference is +0.01
in A(Li).  The star-to-star spread in A(Li) is still dramatic at a factor of
4-5.

There is also a range in A(Li) for the five virtually identical stars in M71.
The Li syntheses for two stars are shown in Figure 3.  These two stars also
show a large difference in the strength of the Li I lines; the determined
abundances are different by a factor of 2.6.  As we found for the M13 stars,
the Li abundances for the M71 stars were very close for the determinations by
the synth and the blends techniques.  The mean difference between the two
methods for the five stars equal to $-$0.03.  The equivalent widths and blends
value for A(Li) are given in Table 4 for the M71 stars, but the determinations
by synth are preferred.

For these stars the value of A(Li) is only dependent on the stellar
temperature and not the other atmospheric parameters.  An uncertainty of
$\pm$50 K corresponds to an uncertainty in A(Li) of $\pm$0.04.  This has no
effect on the {\it relative} Li abundances.  A large change in [Fe/H] of +0.5
could change A(Li) by only +0.01.

\section{DISCUSSION}

The relatively metal-rich cluster, M71, is one of the youngest clusters at
11.2 Gyr, while M13 is more like a classical globular cluster with [Fe/H] =
$-$1.44 and an age of 13.5 Gyr.  These two clusters extend the results for Li
abundances in unevolved stars in the Galaxy's globular cluster system.  We use
the consistently-determined ages of Valcin et al.~(2020) and [Fe/H] values of
Bailin (2019).

Figure 4 shows a compilation of A(Li) in turnoff stars in eight globular
clusters with the references given in the figure caption.  This is an expanded
view of Figure 5 in Boesgaard \& Deliyannis (2023).  For comparison we
show the field-star plateau which is from Charbonnel \& Primas (2005).
This plateau is for unevolved field stars in their ``clean'' sample with
temperatures above 6000 K and [Fe/H] less than -1.5.  (These are important
restrictions on the sample.)  For those field stars, however, there is
also a range in A(Li) of at least 0.5 at a given [Fe/H].  The spread is
also seen in the Li results of Roederer et al.~2014 and Bandyopadhyay et
al.~2022), the latter is focused on Li abundances.  There are also upper
limits on Li abundances found in the field star samples.  There is an
expectation that the more metal-rich of the field stars would have formed from
gas that has been enriched in Li over time by spallation reactions in the
interstellar gas as proposed by Reeves et al.~(1970) and developed by
Meneguzzi et al.~(1971), referred to as galactic cosmic ray spallation (GCR).
Other sources of new Li include novae, supernovae explosions, mass loss from
super-Li-rich AGB stars.  That figure reveals that the Li abundances show a
large spread in each of the eight globular cluster star samples of unevolved
stars.  The range in A(Li) can be a factor of 10 or more.

It is not surprising that {\it none} of the main-sequence field stars has an
upper limit on Li because they have not evolved to where the surface
convection zone has deepened enough to have diluted the Li in the outer
layers.  In the case of the globular cluster stars, however, if they are
second or later generation stars their Li content might be expected to have
been affected by the gas that previous generations of stars spewed into the
newer stars.  That gas would be depleted of some Li during the course of
stellar evolution.  For the globular clusters only NGC6752 has any stars with
Li upper limits.

The conclusions we drew from this same plot for six clusters in our paper
about Li in turn-off stars in M5 by Boesgaard \& Deliyannis (2023) are
strengthened by the addition of these two clusters, M13 and M71.  

The ages of clusters can be determined well through stellar evolution models.
The association of age with A(Li) can be seen In Figure 5.  There does not
seem to be any connection with age other than for the youngest cluster,
M71. It has a lower Li maximum in the five turn-off stars that we
investigated.

For each of the eight clusters there is a large range in A(Li).  The range is
a factor of four for M13 and a factor of 2.4 for M71, while the old but more
metal-rich cluster, 47 Tuc, shows a span of a factor of 10.  Upper limits on
A(Li) have been found only in NGC6752 (Gonzales Hernandez et al.~(2009) and
indicates a range of 10 or more.  For the field stars values for A(Li) cover a
range of about a factor 4 at a given Fe abundance and a factor 10 with over
the full range in [Fe/H] from $-$4 to $-$1 (Deliyannis, Pinsonneault \& Duncan
1993), Bandyopadhyay et al.~2020).  Individual field stars may have formed in
gas that had been enriched by GCR and other mechanisms of Li production.  This
is unlikely to explain the Li range in a given cluster.

We note that some studies of Li in globular clusters have found an
anti-correlation of Na and Li (e.g. for {\underline (only)} four stars in 47
Tuc by Bonifacio et al.~2007).  This could arise from inclusion of
later-generation of stars that might have been formed in part with pristine
material that contains a normal abundance of Li and in part by material
expelled from first generation AGB stars that is devoid of Li and enriched in
Na.  This is {\underline not} the situation for our clusters.  Abundance
results for many elements in M 71 have been published by Boesgaard et
al.~(2005) including Na.  Those Na results and these Li results show that Li
and Na are correlated for these stars in M 71, not anti-correlated.  (We used
4 weak Na lines to find the Na abundances.)  For M13 the Na lines are too weak
to measure because of the lower metal content, [Fe/H] = -1.44.  However, we
took care to select identical, unevolved stars in M13 with V = 17.88 and
(B-V)$_0$ = 0.47.

For the other clusters in our Figure 4 there are two others, in addition
to M71, with no measurements of Na lines: M92 and M5.  Lind et al.~(2009)
looked at Li and Na in 100 stars in NGC6397 and found three with low A(Li),
below 2.0 and high A(Na) above 4.25.  The other 97 stars appear to be randomly
distributed with A(Li) between 2.05 to 2.45 and A(Na) between 3.5 and 4.3.  In
fact, they found no significant anti-correlation for A(Na) values below 4.1.
They suggest that the stars with A(Na) below 3.9 are first generation,
non-polluted stars.  For NGC6752 Gruyters et al.~(2014) show an
anti-correlation of Li and Na for the \underline{low} Li stars.  The
anti-correlation of Li and Na is very weak for M4; Spite et al. (2016) show
that for a span of [Na/Fe] from -0.2 to +0.4 , the change in A(Li) is only
2.22 to 2.10.

Fields et al.~(2020) have made calculations for the nuclei created during Big
Bang nucleosynthesis incorporating the results from the {\it Planck} data.
The source of deuterium ($^2$H) is only primordial; however, it can be
destroyed during stellar and galactic evolution.  That primordial D can be
measured well in high-Z quasar absorption systems.  They use a weighted mean
of 11 D measurements and find primordial D/H = 2.55 ($\pm$0.03) x 10$^{-5}$.
This determines a precise baryon-to-photon ratio in the standard model of Big
Bang Nucleosynthesis, which in turn predicts that A(Li)$_p$ would be near 2.7.
The   value they calculate for Li/H$_p$ is 4.72 $-$ 4.76 x 10$^{-10}$
($\pm$0.74).  This corresponds to A(Li) = 2.67 $-$ 2.68 in A(Li).  This is
shown as the predicted primordial Li in Figures 4 and 5.

In our study of Li in turn-off stars in M92 (Boesgaard et al.~1998) we found a
spread in A(Li) in otherwise similar stars and attributed it to a diversity in
initial angular momenta.  Pinsonneault, Deliyannis \& Demarque (1992) showed
the effects of initial rotation on Li and Li depletion.  They calculated Li
(and Be) values as a function of mass, (turn-off) age, metallicity, rotation,
temperature.  They showed the effects of the rotational evolution of stars,
for the case of metal-poor stars.  Their models are based on the approach of
Endal \& Sofia (1976, 1981) as updated by Pinsonneault et al. (1989).
Population I stars are known to spin up as they contract during the pre-main
sequence, and then spin down as they lose angular momentum during the main
sequence.  In the models the latter creates differential rotation with depth,
which sets off a secular shear instability, which leads to mixing of surface
material with regions where it is hot enough to destroy Li, and thus surface
Li depletion.  Stars forming with a larger initial angular momentum lose more
angular momentum as they evolve, and thus deplete more Li.  So one test of
these models is their prediction that stars of identical T$_{\rm eff}$ should
exhibit different Li abundances due to differences in the initial angular
momenta.

It is important to note that these (and updated) ``Yale''-type models have
passed a number of varied tests in their predictions of the evolution of
surface stellar Li, Be, and B abundances.  Since Li, Be, and B survive to
different depths (temperatures) , knowledge of their ratios provide extremely
sensitive tests of how physical mechanisms act as a function of depth.  So,
for example, the correlated depletion of Li and Be in F dwarfs strongly
support such models that predict rotational mixing (Deliyannis 1994,
Deliyannis \& Pinsonneault 1997) and other similar ones (Charbonnel et
al.~1994) as the physical cause of the Li Dip (Boesgaard \& Tripicco 1986),
while also arguing strongly against other mechanisms proposed to create the Li
Dip such as diffusion, mass loss, and slow mixing due to gravity waves
(Deliyannis et al.~1998, Boesgaard et al.~2004).  Other example tests of the
``Yale''-type models include the correlated depletion of Be and B observed in
F dwarfs (Boesgaard et al.~2005, 2016); the main sequence Li depletion in G
dwarfs (Boesgaard et al.~2022, Sun et al.~2023); the higher than average A(Li)
seen in short period tidally locked binaries (Thorburn et al.~1993, Deliyannis
et al.~1994, Ryan \& Deliyannis 1995); the correlation between Li depletion
and spindown in late A/early F dwarfs (Deliyannis et al.~2019); and the
declining Li abundances with lower Teff in M67 subgiants evolving out of the
Li Dip whose deepening surface convection zones reveal the Li profile
(dependence of Li with depth) created during the main sequence, and especially
the Li/Be ration in these stars (Sills \& Deliyannis 2000; Boesgaard et
al.~2020).  The depletion of Li in late G and K dwarfs may require the
additional ingredient of magnetically-induced radius inflation (Somers \&
Pinsonneault 2015; Jeffries et al.~2021).  It is also worth noting that
``Yale''-type models have enough core rotation to explain rapidly-rotating
horizontal branch stars while also maintaining very slow surface rotation of
these stars main sequence progenitors (Pinsonneault et al.~1991).

Given these and other successes of the ``Yale''-type models, it is reasonable
to infer that these types of models should also apply to halo dwarfs, so that
Li differences might be reasonably tied to differences in the initial angular
momenta.  However, it is more challenging to infer what the predicted Li
depletion should be because we lack key information about halo stars, all of
which are old.  Such information is empirically determined in Population I
stars: from very young stars we know the initial angular momenta as a function
of mass, and from open clusters of different ages we also know how stars of
different masses lose angular momentum over time.  So, for example, some
``Yale'' models predict Li depletion by 1 dex (Pinsonneault et al.~1992,
Chaboyer \& Demarque 1994) whereas others with different assumptions about the
missing information predict as little depletion as 0.3-0.5 dex (Deliyannis
1990, Pinsonneault at el.~1999).  So while a precise prediction of Li
depletion in halo dwarfs from ``Yale'' models is not yet possible, in the
context of such models a depletion of $\sim$0.5 dex from a Big Bang value of
A(Li) $\sim$2.7 to the observed Spite plateau near $\sim$2.2, including
significant Li dispersion, is at least quite reasonable.

All of our observed stars are now now rotating slowly, in fact so slowly that
we can not measure v sin i from our spectra.  The stars with the highest
initial angular momentum would destroy more Li as they spin down while those
with low initial angular momentum would be able to preserve more Li.  The
range in initial angular momenta would result in the observed range in Li.
Slow rotation would allow atmospheric diffusion of Li to occur and that
could play a role in decreasing the Li abundance (Deliyannis 1990, Grutyers,
Nordlander \& Korn 2014).

These unevolved cluster stars all have Li detections with the exception
of a few in NGC 6752.  This seems to put restrictions on the idea of multiple
generations of star formation.  Stars formed later from material that has
passed through the giant phase would have had its Li diluted; new stars would
have reduced or undetectable Li.

We have noticed that the sample size in most of the these clusters is
is only a few, less than ten, yet a range in Li abundance is found in these
otherwise identical unevolved stars. 
 
\section{SUMMARY AND CONCLUSIONS}

We were able to obtain high-resolution spectra of unevolved stars in two
additional globular clusters, M13 and M71.  One major purpose was to ascertain
the Li content which is thought to reflect the primordial Li formed during the
first 15 minutes of the Big Bang.  The stars observed for this project in
those two clusters and have V magnitudes of 17.6 to 17.9 and required
multi-hour exposures of the Keck I telescope with HIRES.  The four stars we
chose to observe in M 13 are virtually identical to each other with V at 17.9
and (B-V)$_0$ of 0.47.  The same is true of the of the five stars in M71 which
have V at 17.7 and (B-V)$_0$ of 0.53.  

 We have determined their Li abundances with spectrum synthesis techniques.
The Li abundances are sensitive to temperature where an uncertainty of $\pm$50
K gives an uncertainty of $\pm$0.04 in A(Li).  As can be seen in Figures 2 and
3 the Li lines can be weak and there is uncertainty in the fit resulting from
that and from the S/N ratio of the original data.  Even so, those figures also
show a clear range in the Li line strengths and abundances.

Main sequence and turn-off stars have been measured for Li abundances in eight
globular clusters and we have shown those Li results in Figure 4 with Fe
abundance where [Fe/H] ranges from $-$2.24 for M92 to $-$0.75 for 47 Tuc
(Bailin 2019).  It is very striking that the unevolved stars in all eight
clusters, including M13 and M71, and also unevolved field stars, show a wide
range in A(Li).  The range can be explained by differences in the initial
angular momentum of the individual stars with the most rapid rotators losing
more of their initial Li during the spin-down to the main sequence, thereby
creating a dispersion in Li abundances.

All of the clusters have stars with Li abundances above those found in the
field-star Li plateau.  And the clusters all have stars with Li below that
plateau.  Only one cluster, NGC 6752, has any unevolved stars with no
detectable Li.

The {\it maximum} amount of Li in each cluster -- in these unevolved stars --
is very similar at A(Li) = 2.5.  This amount of Li, however, is less than the
predictions for the amount of primordial Li produced during nucleosynthesis
during the Big Bang of 2.72 according to Fields et al.~(2020).  This has been
called ``The Lithium Problem.''  Although this problem is now smaller, a
factor of 1.66, it still persists.  

We also show the Li abundances with cluster age in these unevolved stars in
the same eight clusters in Figure 5.  The ages ranges from 14.21 Gyr for NGC
6397 to 11.21 Gyr for M71 according to Valcin et al.~(2020).  The ages of
globular clusters can be well-estimated with stellar evolution models as
opposed to field stars where there can be a large age spread at a given
metallicity.  The samples of stars studied in each of these clusters have a
uniform age as first-generation stars, and presumably a uniform initial
composition.
 
We can reasonably attribute the spread in Li content in stars that are
otherwise virtually identical in each cluster to differences in the initial
angular momentum of the protostars in each star-forming proto-cluster.  During
the stellar spin-down the surface Li is mixed to deeper layers and destroyed
by nuclear reactions with a proton at a temperature of 2.5 x 10$^6$ K.  All
the stars are now slow rotators but those with the greatest initial rotation
have had the largest reduction in initial Li. Effects such as atmospheric
diffusion may also contribute to the observed spread in Li abundances.

\acknowledgments We wish to express our appreciation the Keck Observatory
support astronomers for their assistance and knowledge during our observing
runs.  CPD acknowledges support through the National Science Foundation
grant AST-1909456.

Facility: Keck I HIRES Software; IRAF (Tody 1986, 1993);MOOG (Sneden 1973;
Sneden et al. 2012)

 ORCID IDs Ann Merchant Boesgaard https://orcid.org/0000-0002-8468-9532
Constantine P. Deliyannis https://orcid.org/0000-0002-3854-050X

\clearpage
\tablenum{1}
\begin{center}
\begin{deluxetable}{rccclccc} 
\tablewidth{0pc}
\tablecolumns{8} 
\tablecaption{Log of the Keck/HIRES Observations of M13 Turn-Off Stars} 
\tablehead{ 
\colhead{Star (PS)}  &  \colhead{V}  & \colhead{B-V}  
& \colhead{(B-V)$_{\rm o}$} 
& \colhead{Date (UT)}
& \colhead{Exp(min)} 
& \colhead{Total(min)} 
& \colhead{S/N}  
}
\startdata 
33857\phn   & 17.884 & 0.513 & 0.470 & 1997 May 27 & 2$\times$45 &  & \\ 
            &        &       &       & 1998 Jun 21 & 3$\times$45 & 225 & 45 \\
38266\phn   & 17.885 & 0.513 & 0.470 & 1997 May 27 & 4$\times$45 &  & \\
            &        &       &       & 1998 Jun 21 & 4$\times$45 & 360 & 65 \\
45212\phn  & 17.880 & 0.512 & 0.463 & 1997 May 27 & 3$\times$45 &  & \\
            &        &       &       & 1997 Aug 30 & 3$\times$45 &  & \\
            &        &       &       & 1998 Jun 21 & 2$\times$45 &  & \\
            &        &       &       & 1998 Jun 22 & 4$\times$45 & 540 & 72 \\
45588\phn   & 17.880 & 0.518 & 0.475 & 1998 Jun 22 & 3$\times$45 & 135 & 36 \\
\enddata 
\end{deluxetable} 
\end{center}

\begin{deluxetable}{rccclccc}
\tablenum{2}
\tablewidth{0pc}
\tablecaption{Log of the Keck/HIRES Observations of M71 Turn-Off Stars}
\tablehead{
\colhead{Star} & \colhead{V} & \colhead{B-V} & \colhead{(B-V)$_0$} 
& \colhead{Date(UT)} & \colhead{Exp(min)} &
\colhead{Total(min)}  &\colhead{S/N} 
} 
\startdata 
239  & 17.62  & 0.81 & 0.53  & 1997 Aug 31 & 2$\times$45 &     &  \\
     &        &      &       & 1998 Sep 10 & 1$\times$60 &     &  \\
     &        &      &       & 1998 Sep 11 & 1$\times$45 & 195 & 60 \\
259  & 17.66  & 0.82 & 0.54  & 1996 Jul 26 & 1$\times$60 &     & \\
     &        &      &       & 1996 Jul 27 & 1$\times$45 &     & \\            
     &        &      &       & 1997 Aug 30 & 2$\times$45 &     & \\
     &        &      &       & 1998 Sep 11 & 1$\times$45 & 240 & 50 \\
260  & 17.66  & 0.79 & 0.51  & 1997 Aug 31 & 2$\times$45 &     &  \\
     &        &      &       & 1997 Aug 31 & 1$\times$30 &     &  \\
     &        &      &       & 1998 Sep 11 & 1$\times$60 & 180 & 50 \\
264  & 17.67  & 0.81 & 0.53  & 1997 Aug 31 & 3$\times$45 & 135 & 48 \\    
273  & 17.68  & 0.81 & 0.53  & 1996 Jul 27 & 1$\times$45 &     & \\
     &        &      &       & 1997 Aug 30 & 2$\times$45 &     & \\
     &        &      &       & 1998 Jun 23 & 1$\times$45 &     & \\
     &        &      &       & 1998 Sep 10 & 1$\times$60 &     & \\
     &        &      &       & 1998 Sep 10 & 1$\times$45 & 225 & 60 \\
\enddata
\end{deluxetable}

\clearpage
\singlespace
\begin{center}
\begin{deluxetable}{lccccccc} 
\tablewidth{0pc}
\tablenum{3}
\tablecolumns{5} 
\tablecaption{Model Parameters and Abundances for M13 Stars} 
\tablehead{ 
\colhead{Star}  &  \colhead{Temperature}  & \colhead{$\log{\rm g}$}  &
\colhead{$\xi$} & \colhead{[Fe/H]} & \colhead{EQW(Li)} & \colhead{A(Li)} 
& \colhead{A(Li)} \\
\colhead{}	&  \colhead{(K)} & \colhead{(dex)}  & \colhead{(km~s$^{-1}$)}  
  &  \colhead{(dex)} &  \colhead{(m\AA{})} &  \colhead{(blends)} & 
\colhead{(synth)}
}
\startdata 
33857.........  & 5890$\pm45$ & 3.80 & 1.50 & $-$1.44 & 14.2 &1.79 & 1.85 \\

38266.........  & 5900$\pm45$ & 3.80 & 1.50 & $-$1.44 & 50.5 &2.43 & 2.45  \\

45212.........  & 5840$\pm45$ & 3.80 & 1.50 & $-$1.44 & 21.0 &1.94 & 2.01  \\

45588.........  & 5910$\pm45$ & 3.80 & 1.50 & $-$1.44 & 56.0 &2.49 & 2.38 \\

\enddata 
\end{deluxetable} 
\end{center}


\singlespace
\begin{center}
\begin{deluxetable}{lccccccc} 
\tablewidth{0pc}
\tablenum{4}
\tablecaption{Model Parameters and Abundances for M71 Stars}
\tablehead{
\colhead{Star} & \colhead{Temperature} & \colhead{$\log{\rm g}$} & 
\colhead{$\xi$} & \colhead{[Fe/H]} & \colhead{EQW(Li)} & \colhead{A(Li)} 
& \colhead{A(Li)} \\
\colhead{} & \colhead{(K)} & \colhead{(dex)} & \colhead{(km~s$^{-1}$)} 
& \colhead{(dex)} & \colhead{(m\AA{})}  & \colhead{(blends)} &
\colhead{(synth)} 
}    
\startdata
239.........  & 5845$\pm45$ & 4.01 & 1.00 & $-$0.80 & 30.0 & 2.12 & 2.20 \\
259.........  & 5800$\pm45$ & 3.98 & 1.00 & $-$0.80 & 28.1 & 2.05 & 2.10 \\
260.........  & 5930$\pm45$ & 4.08 & 1.00 & $-$0.80 & 43.3 & 2.38 & 2.28 \\
264.........  & 5845$\pm45$ & 4.01 & 1.00 & $-$0.80 & 29.5 & 2.11 & 2.05 \\
273.......... & 5845$\pm45$ & 4.01 & 1.00 & $-$0.80 & 24.0 & 2.01 & 1.87 \\
\enddata 
\end{deluxetable} 
\end{center}


\clearpage

\begin{figure}
\plottwo{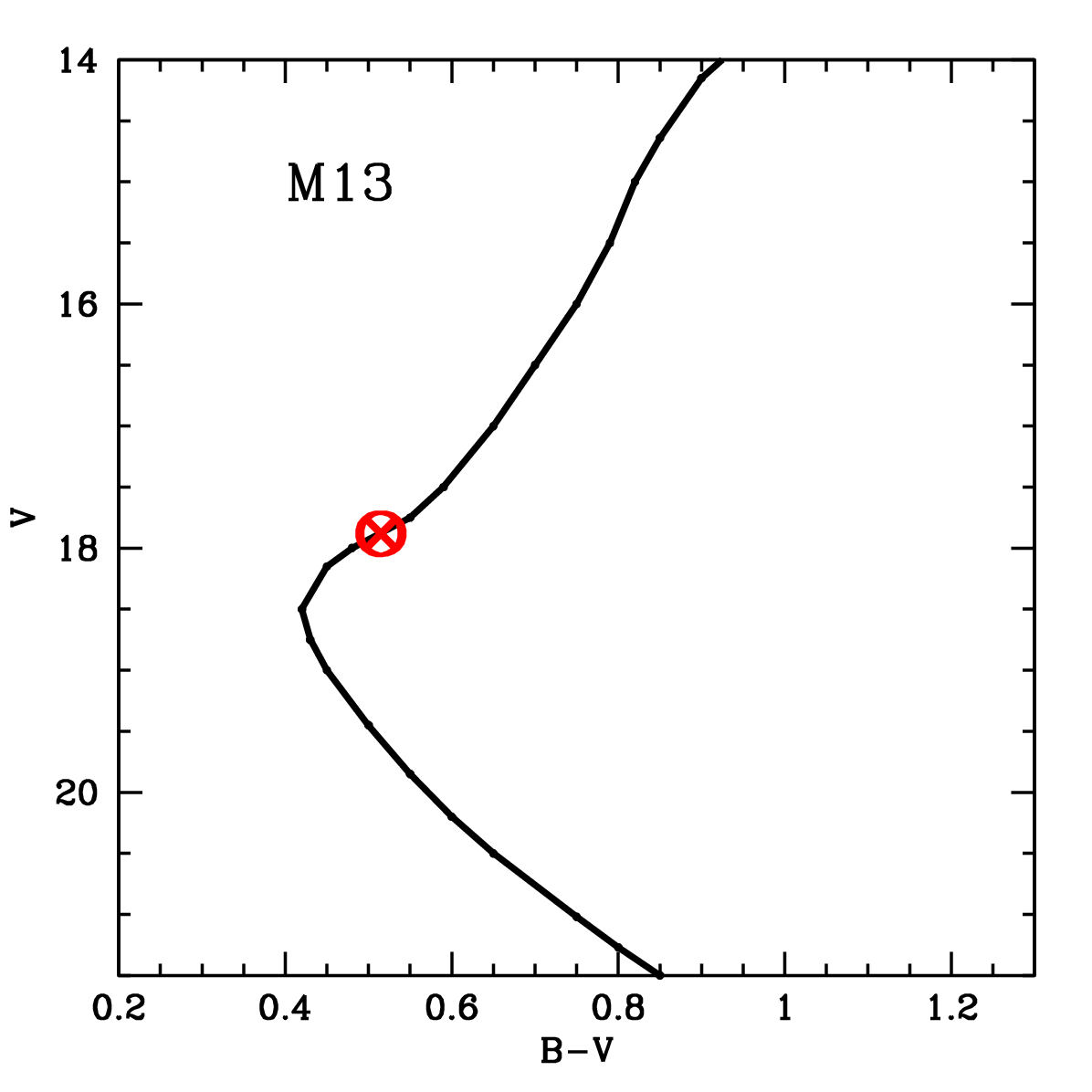}{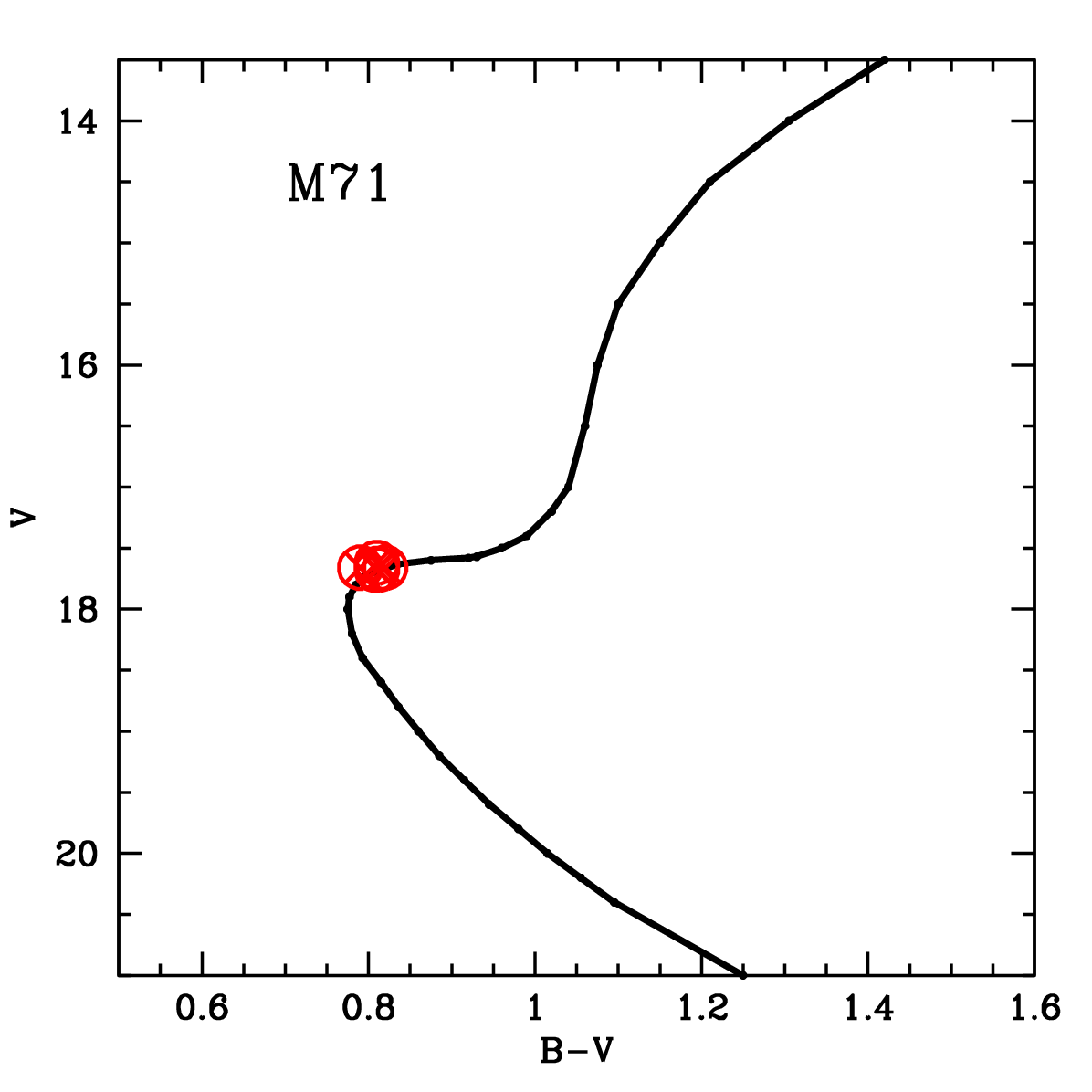}
\caption{Left: Color-magnitude diagram for M13 with the fiducial sequence from
Sandage (1970).  The four stars we observed are shown as red circled X
symbols; note their similarity.  Right: Color-magnitude diagram for M71 with
the fiducial sequence from Hodder et al.~(1992).  Our five nearly-identical
 observed stars are shown as red circled X symbols.  The two plots are on the
same vertical and horizontal scales.  Note the faintness of the stars, near V
= 18.}
\end{figure}

\begin{figure}
\plotone{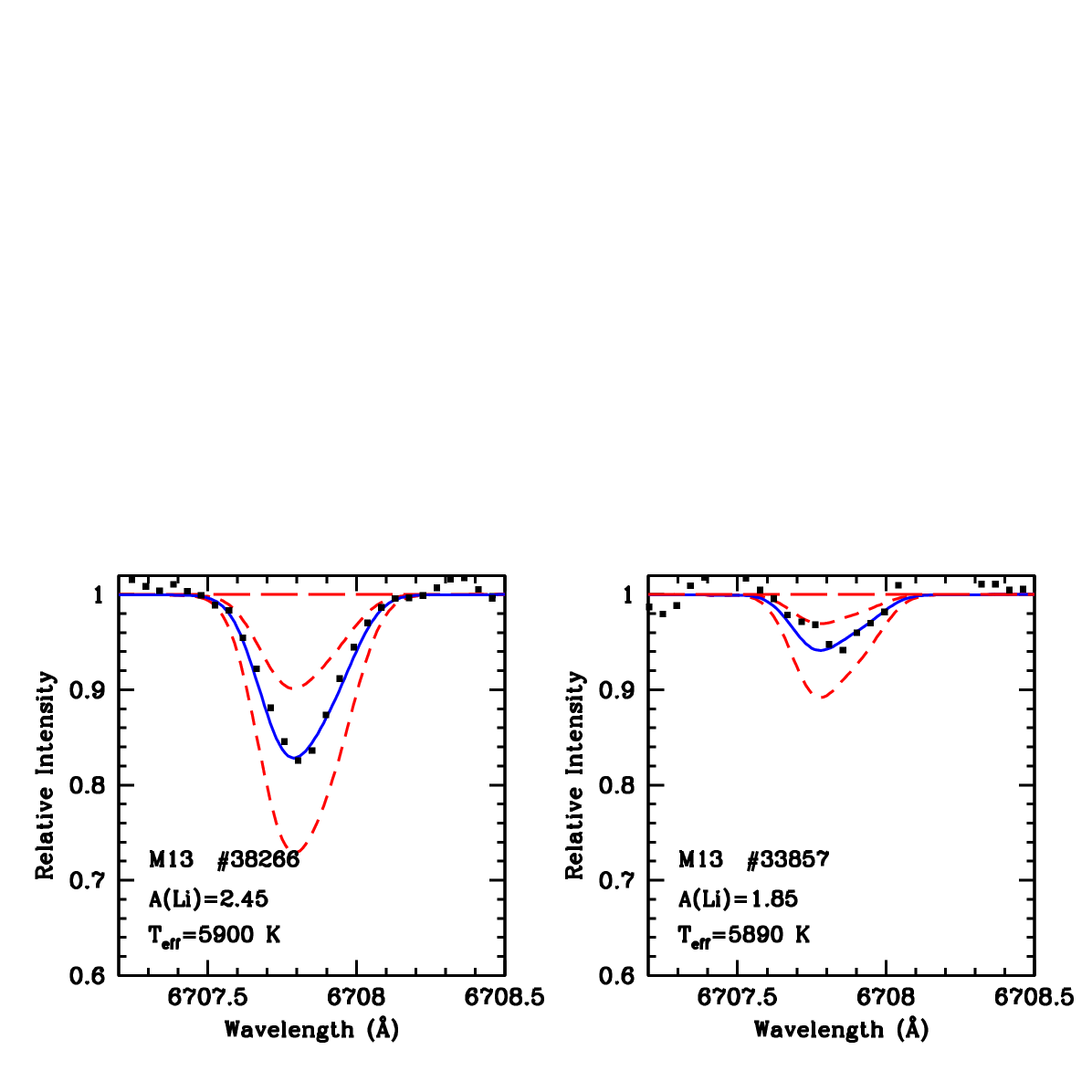}
\caption{The spectrum synthesis for two of our four M13 stars.  The star on
the left, with the strong Li I line, is 38266, while the star on the right
with the weaker Li line is 33857.  The observed points are the black squares.
The solid blues line is the best fit synthesis.  The red dashed lines are a
factor of two more and two less Li, while the long-dashed red line represents
no Li at all.  These two stars differ in temperature by only 10 K, but their
Li content differs by a factor of four.}
\end{figure}

\clearpage

\begin{figure}
\plotone{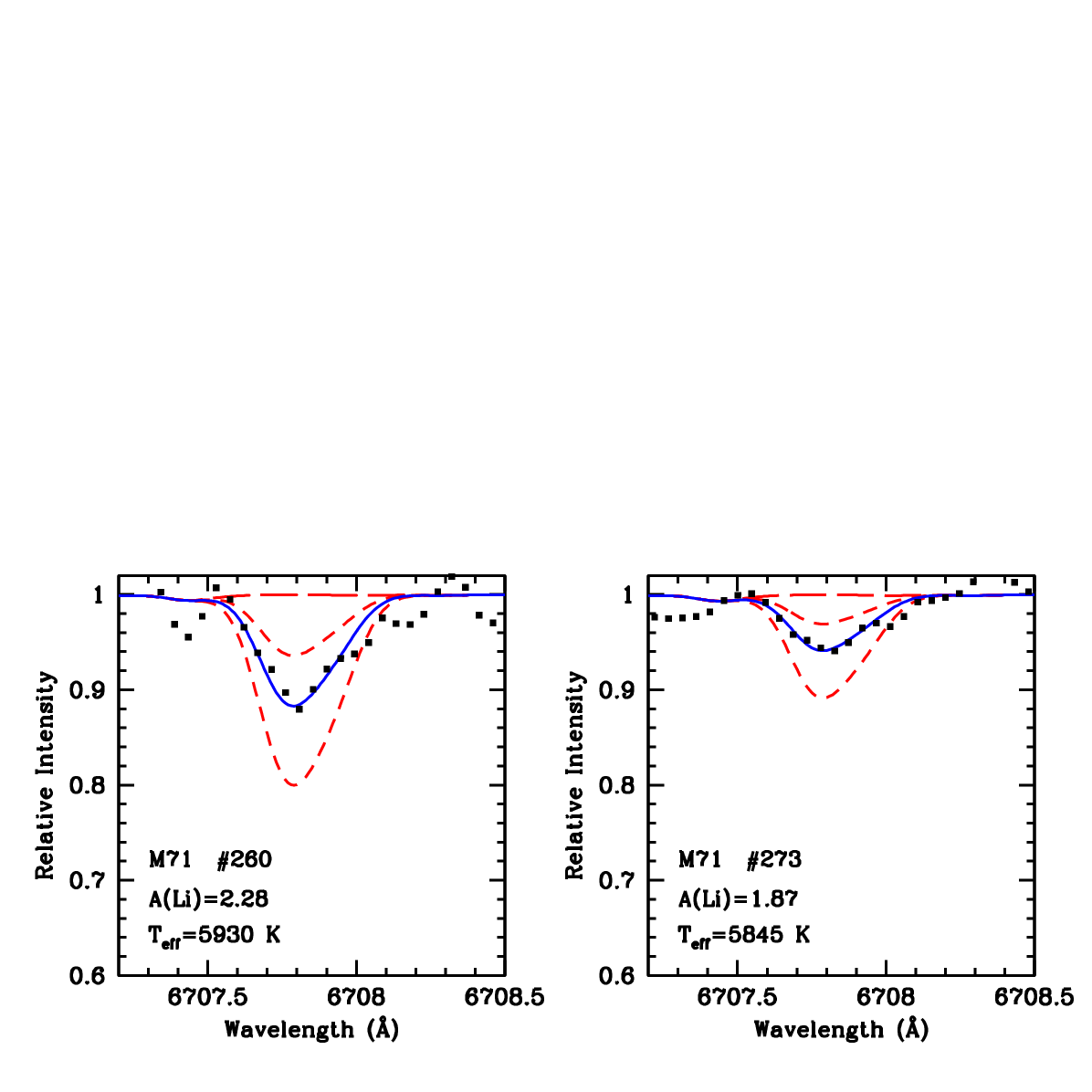}
\caption{The spectrum synthesis for two of our five M71 stars.  The stars on
the left is \#260 with a Li content A(Li) of 2.28 which is a factor of 2.6
higher than that of \#273 at A(Li) = 1.87.  The lines and symbols are the same
as in Figure 2.}
\end{figure}

\clearpage

\begin{figure}
\epsscale{0.6}
\plotone{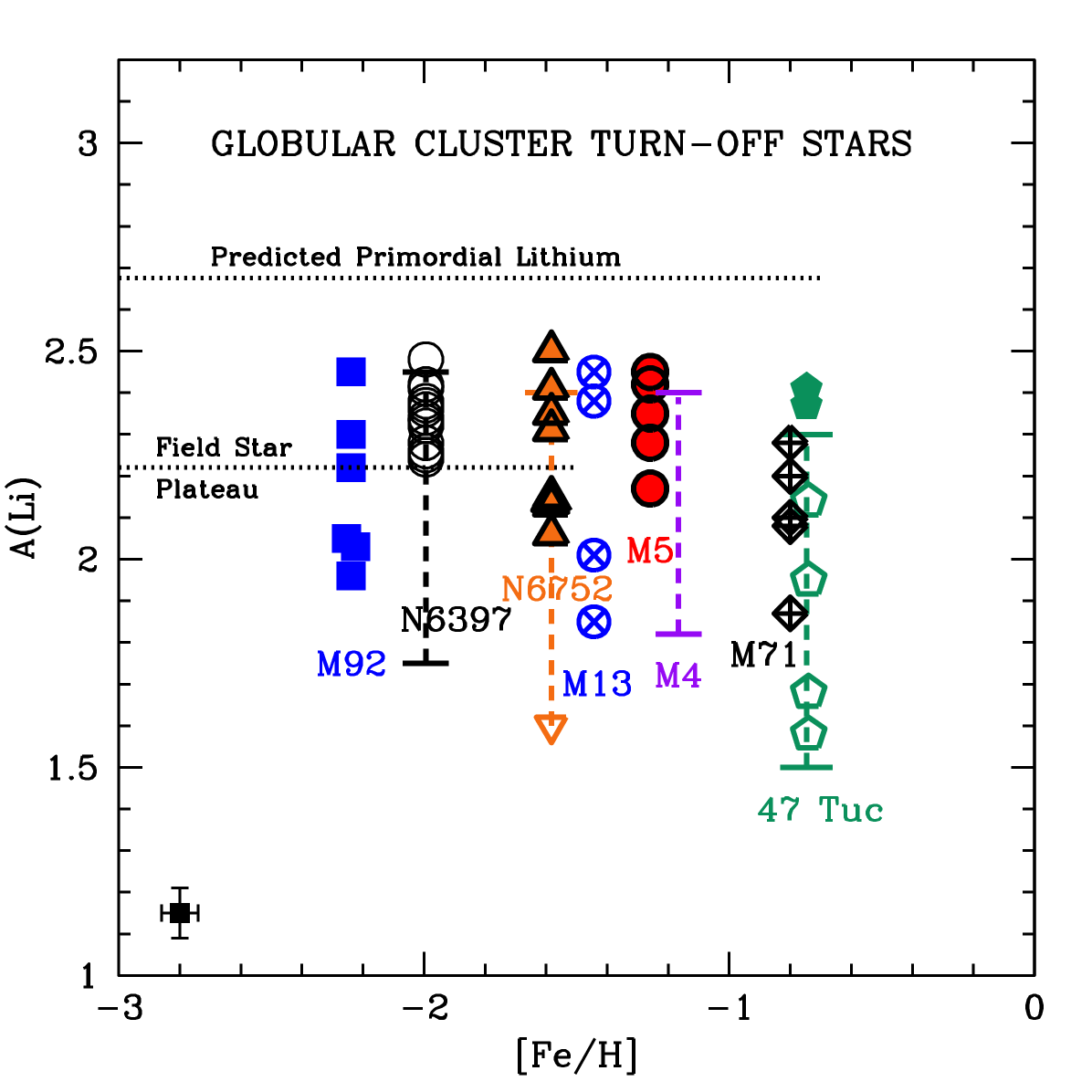}
\caption{The Li abundances in unevolved stars in eight globular clusters
according to their Fe abundances.  The results from this work for four stars
in M13 are shown as black-circled crosses and those for M71 as black-squared
plus signs.  A typical error bar for our work is shown in the lower left
corner.  Our results for M92 (Boesgaard et al.~1998) are blue-filled squares,
with two stars with identical values are offset for clarity.  The open black
circles are for NGC 6397 are from Pasquini \& Molaro (1996), Nordandler
et al.~(2012), Lind et al.~(2009) and Bonifacio et al.~(2002).  Additional
results for NGC 6397 are indicated by the vertical dashed black line are from
Gonzales Hernandez et al.~(2009).  For NGC 6752 the orange triangles are from
Pasquini et al.~(2005) while the vertical orange-dashed line shows the range
for turn-off stars hotter than 5850 K from Gruyters et al.~(2014) and
Schiappacasse-Ulloa et al.~(2022).  For M5 the filled red circles are from
Boesgaard \& Deliyannis (2023) for five turn-off stars.  The range shown for
the Li results for M4 by the vertical purple dashed line are from Monaco et
al.~(2012) and Spite et al.~(2016).  Pasquini \& Molaro (1997) determined Li
in 2 stars in 47 Tuc, shown as filled green pentagons and and those by
Bonifacio et al.~(2007) are filled green pentagons.  The major study by Aoki
et al.~(2021) included 93 turn-off stars is represented here by the vertical
dashed green line.  The horizontal line at A(Li) = 2.22 represents the
plateau for unevolved, metal-poor stars ([Fe/H] $<$-1.5) with temperature
greater than 6000 K from the clean sample of Charbonnel \& Primas (2005).
The diagonal red-dashed line labeled ``Dwarfs B+'' is the upper envelope for
Li in field star dwarfs of Bandyopadhyay et al.~(2022), based on Li data
from Roederer et al.~(2014).  The horizontal line at A(Li) = 2.719 represents
the prediction for primordial Li from Big Bang Nucleosynthesis including
results from Planck and WMAP from Fields et al.~(2020).}
\end{figure}

\clearpage

\begin{figure}
\epsscale{1.0}
\plottwo{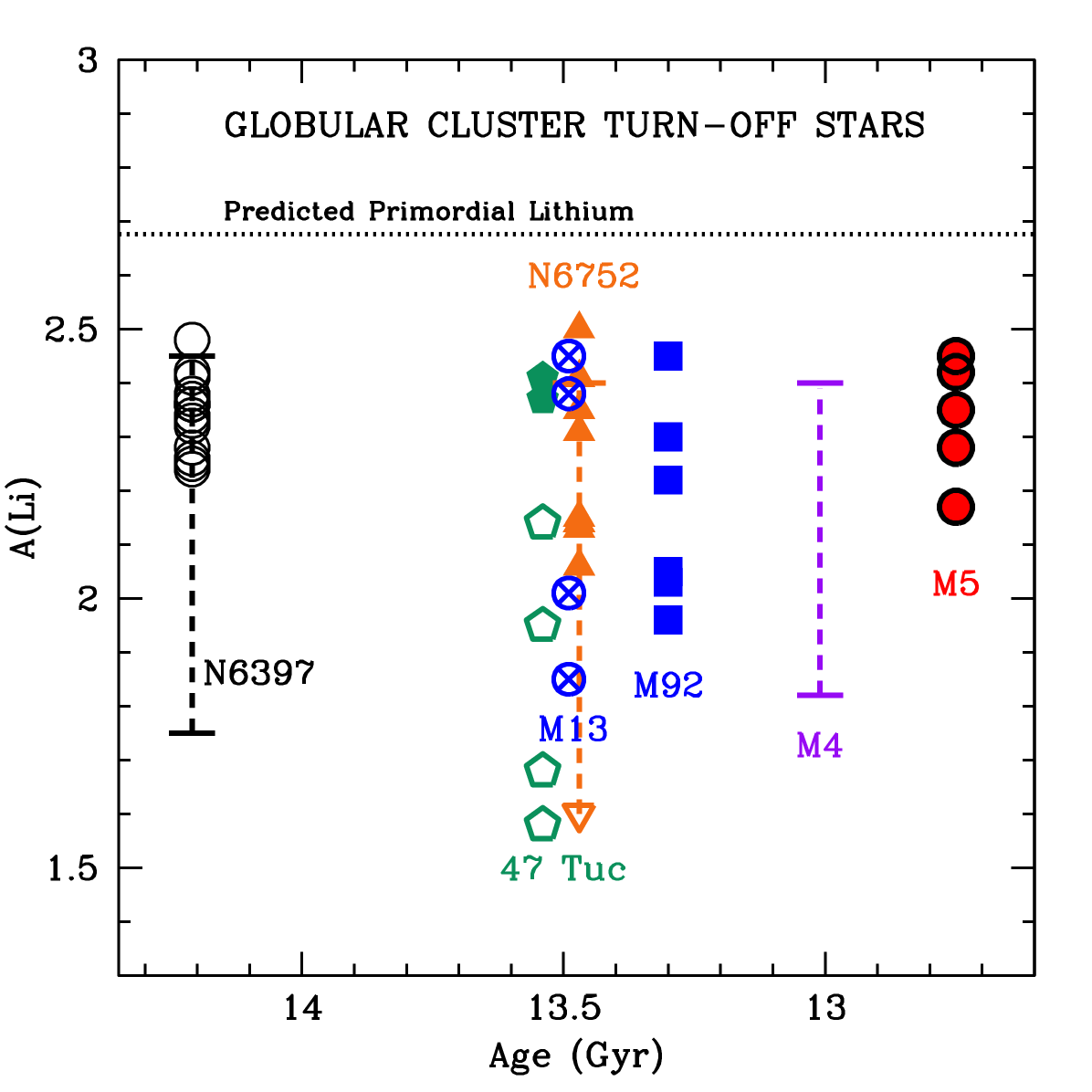}{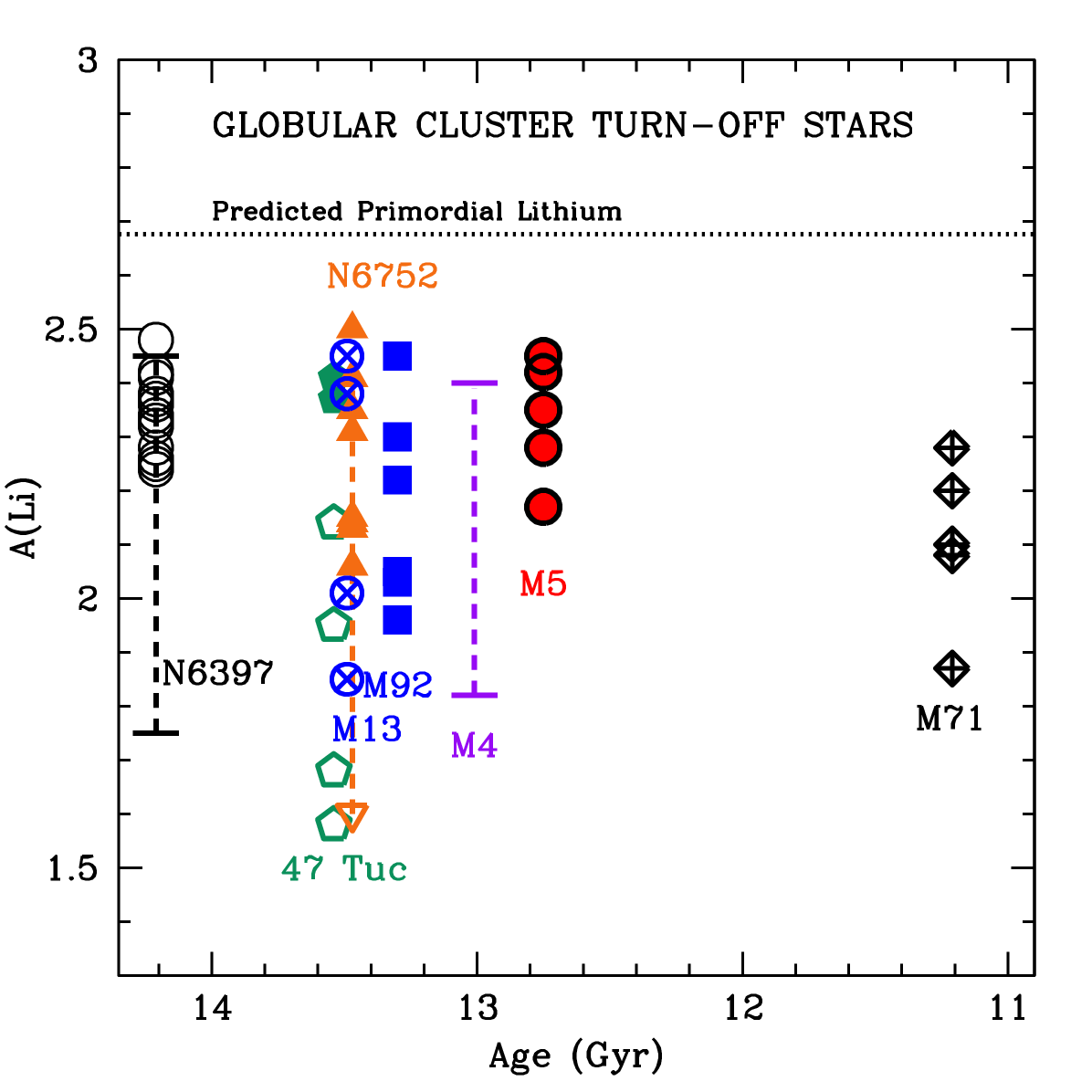}
\caption{Left: The counterpart to Figure 4 with cluster age as the abscissa. 
 The ages are from Valcin et al.~(2020).  Right: The x-axis is expanded to
include the youngest cluster, M71, at 11.21 Gyr.}  
\end{figure} 

\end{document}